\documentclass[review]{elsarticle}

\journal{Journal of Physica A}

\begin{document}

\begin{frontmatter}

\title{Order-Disorder structural transition in a confined fluid}

\author{E. M. de la Calleja-Mora}
\address{Instituto de F\'{i}sica, Universidade Federal do Rio
Grande do Sul, Caixa Postal 15051, 91501-970, Porto Alegre, RS, Brazil}

\author{Leandro B. Krott}
\address{Campus Ararangu\'a, Universidade Federal de Santa Catarina, Rua
Pedro Jo\~ao Pereira, 150, CEP 88900-000, Ararangu\'a, SC, Brazil}

\author{M. C. Barbosa}
\address{Instituto de F\'{i}sica, Universidade Federal do Rio
Grande do Sul, Caixa Postal 15051, 91501-970, Porto Alegre, RS, Brazil}

\begin{abstract}
In this paper the amorphous/solid to disorder liquid structural phase transitions of an anomalous confined fluid is analyzed using their local fractal dimension. The model is a system of particles interacting through a two length scales potentials confined by two infinite plates. In the bulk, this fluid exhibit water-like anomalies and under confinement forms layers of particles. The particle distributions of them, present different arrangements related to amorphous/solid phases. Here only the contact layer is analyzed through fractal singularity spectrum. At high densities the structural transition its quantify by the order degree to determine the phases affected by the confinement. This mapping shows that the system as the temperature increased, the fractal dimension decreases, which is consistent with the behavior studying in such systems. This result suggests that under thermodynamic perturbations, an anomalous confined liquid, presents different phase transitions achieving be characterized by its
fractality.
\end{abstract}


\end{frontmatter}

%\linenumbers

%%%%%%%%%%%%%%%%%%%%%%%%%%%%%%%%%%%%%%%%%%%%%
\section{Introduction}
%%%%%%%%%%%%%%%%%%%%%%%%%%%%%%%%%%%%%%%%%%%%%

The characterization of the  phases present in complex system
is not trivial. Usually it depends in identifying the correct order
parameter of the structure.
For instance, the structural transformation by thermal or mechanical perturbation
of fibrous, dendritic, or colloidal configurations, formed by aggregation or
reaction processes have been quantified from different measures of complexity.
One of these measures is the fractal dimension. The fractality is a
geometrical, topological and structural property present in many natural, physics
or simulated complex systems ~\cite{Vicsek,Barnsley,Benoit}. In many cases
a  fractal structure results from the kinetic aggregation of a group of particles
or from the reaction processes between them ~\cite{Vicsek,Carrillo,Meakin1} in
a process that resembles a very slow nucleation and growth of mechanism. This is
visually manifested by different final distributions of the particles that
entails universal properties, and also influence the physical, biological or chemical properties on the system~\cite{Carrillo,Meakin1,Meakin,Blair,Matsushita,Ben-Jacob}. These distributions can be also quantified by the mass fractal dimension that is a measure employed to quantify the different structural phase transition~\cite{Weitz,Witten,Snezhko}.

In a number of systems a single fractal dimension is unable to capture the
full complexity of the system.
The multi-fractal spectrum describes the scaling correlations, coexisting in
the dynamical evolution of the system, at different length scales of
observation. It is employed to provide
a description of the aggregation kinetics. It also gives the
information about how the new phase reaches the equilibrium
state~\cite{Benoit,Bacry,Jensen}. In this way the measure of the
microscopic multi-scale structure through the local fractal dimension, is
 an important tool to identify the macroscopic state of the system, which
influence the physical emergent properties.

Then, the fractal dimensions can be used as
an additional tool for
characterizing the complex phases emerging
from phase transitions. This strategy was employed in the study of
rheological fluids~\cite{Calleja1,Calleja2}, granular
materials~\cite{Blair,Gonzales,Pusey}, magnetic wall domains
in boracite~\cite{Carrillo2} and other
complex  systems~\cite{Witten,Suzuki,Gonzalez2}.
In the case of  rheological systems, the final structures obtained
 by magnetic particles dispersed in mineral oils and perturbed
by magnetic fields show different degrees of order
that were  quantified by its mass fractal dimension~\cite{Calleja1}.
This result was also checked experimentally~\cite{Calleja1,Gonzales}. This
structural transformations can be analyzed by glass transition
approach~\cite{Debenedetti1,Shell,Stillinger3} and is possible to identify the
liquid-glass and the liquid-crystal phase transition.

All the examples cited above in which the fractal analysis were
used to identify new phases were complex systems. Would
this framework also be useful for describing phases
in simple systems?
In a simple fluid the thermodynamic and dynamic behavior is governed
by the molecular length scale. This is the case of the rare gases, diatomic
and triatomic molecules. In a complex fluid, the thermodynamic and
dynamic properties are governed not by the atomistic length scales
but by a mesoscopic scale that arises
from the competition of the multi-scale molecular forces. These
systems include colloidal suspensions, gels and polymer blends. Due to
the complexity of the
competition forces, complex fluids can be considered homogeneous at the
macroscopic scale, but are disordered at the microscopic scale, and
possess structure at an intermediate scale. This is
the reason why  the
multi-fractal spectrum employed to analyze the structural
transformation by mechanical perturbations can be applied in
those complex fluids as well.

Water, even though a very simple triatomic molecule, is not a
simple liquid. It is an anomalous material showing a number of
thermodynamic and dynamic anomalies~\cite{URL}. The most familiar
anomaly is its increasing density with temperature, at ambient
pressure, up to  $4^{\rm o}{\rm C}$. Above
this temperature water behaves as a normal liquid and density
decreases as temperature rises. Experiments for water allow to
locate the line of temperatures of maximum density (TMD) in the
pressure-temperature plane. Below TMD, density decreases with
decreasing temperature, differently from the behavior of the
majority of fluids, for which density increases on lowering
temperature \cite{An76}.

In addition  to the thermodynamic and dynamic anomalies, water
exhibits many solid phases. Several coexistence lines separate
the multiple solid phases. Thus, the energy landscape associated
to the crystalline phases presents a number of sharp valleys with
very low energies. The temperature and pressure ranges at which each
one of these sharp valleys displays lowest energy values define the
stable phase in that region of the phase diagram. Those valleys of the
 energy landscape that never achieve the lowest energy correspond to
the amorphous configurations. When confined within plates, the fluid
energy landscape becomes even more complex. The anomalous fluid forms
layers and the system shows a transition from three layers to a two
layers structure~\cite{Bordin14a,Gi09,KrB15a}. Using
nanotubes, the same transition appears and it is associated with a
dynamic transition from a normal to superflow
regime~\cite{Jakobtorweihen05,Qin11,Bordin13a,Bordin14b}. At low
 temperatures and high degree
of confinement the contact layers melt, while the central layer
stays liquid. The contact layer form a variety of liquid, liquid
crystal and solid structures~\cite{Bordin14c}.

Recently a model for describing
the anomalous behavior of water were studied under
confinement~\cite{Bordin14a}. These studies
indicate that the confined system exhibits at the
wall  two dimensional phases
not present in the bulk system~\cite{Gi09,Bordin14c,Er01,ryzhov14a,BoK15a}.
While the existence of the phases is identified
clearly by the
instabilities of the density versus pressure phase diagram,
the nature of the new structures, tested
with the radial distribution function~\cite{Bordin14a} and  with
the translational order parameter
~\cite{Gi09,Bordin14c,Er01,ryzhov14a,BoK15a}, it is still
unclear. Particularly the system presents phases that change
continuously to very different structures without phase transition
while other phases change through a first order transition.
These two scenarios can not be identified by the translational order parameter
analysis and need further understanding.

In this work we explore the idea that the fractal analysis can
provide information of the structure and phase behavior
of anomalous fluids, like water. In this context
we study the phase behavior of a water-like model
confined within plates. The pressure versus temperature phase diagram, of
this fluid is  analyzed in the framework of the multi-fractal
spectrum and within this framework the different phases
are identified.

The paper is organized as follows: in the
section II the model is introduced; in the section III
the methods are presented;   the results are given in the section IV and
our final
conclusions are presented  in the section V.

%%%%%%%%%%%%%%%%%%%%%%%%%%%%%%%
\section{The Model}
%%%%%%%%%%%%%%%%%%%%%%%%%%%%%%%%%

\begin{figure}[!htb]
\begin{center}
\includegraphics[width=8cm]{fig1a.eps}
\includegraphics[width=8cm]{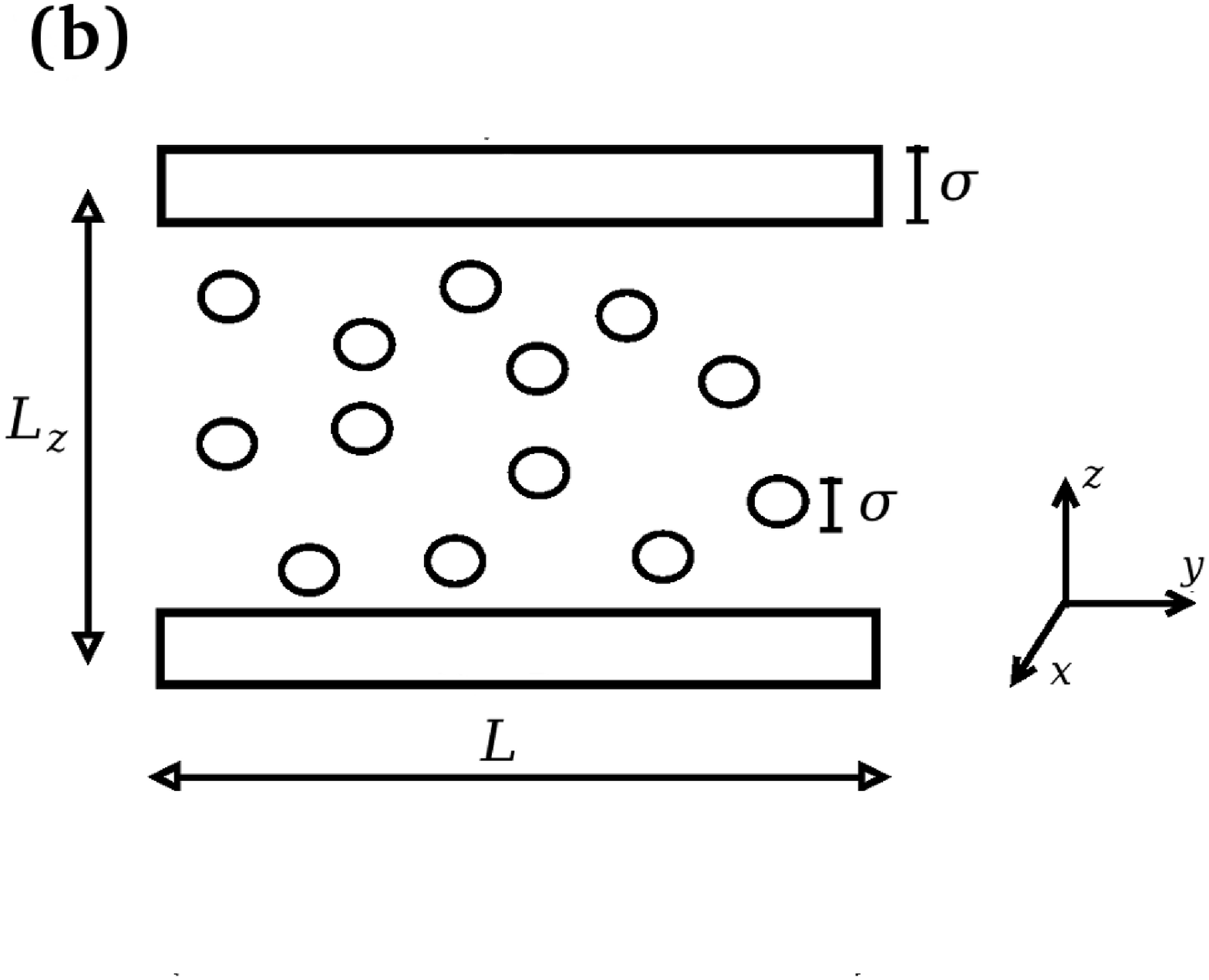}
\end{center}
\caption{(a) Two length scale interaction potential and (b)
schematic depiction of the confined
system. The particles are confined between two flat and smooth
plates, separated by a
distance $L_z$. The energy and length scales are given by
$\epsilon$ and $\sigma$, respectively.}
\label{fig1}
\end{figure}

The water-like fluid is composed by $N$ spherical particles of effective
diameter $\sigma$ that interact through a core-softened potential
of two length scales, namely
%%%%%%%%%%%%%%%%%%%%%%%%%%%%
\begin{equation}
\frac{U(r_{ij})}{\epsilon} = 4\left[ \left(\frac{\sigma}{r_{ij}}\right)^{12} -
\left(\frac{\sigma}{r_{ij}}\right)^6 \right] + a {\rm{exp}}\left[-\frac{1}{c_0^2}\left(\frac{r_{ij}-r_0}{\sigma}\right)^2\right]
\label{AlanEq}
\end{equation}
%%%%%%%%%%%%%%%%%%%%%%%%%%%%
\noindent where $r_{ij} = |\vec r_i - \vec r_j|$ is the distance
between two fluid particles $i$ and $j$. The first term is a standard
12-6 Lennard-Jones (LJ) potential~\cite{Allen} and the second one is a
Gaussian well centered at $r_0$, with depth $a$ and width $c_0$. The
parameters used in this work are $a = 5.0$, $c_0 = 1.0$ and
$r_0/\sigma = 0.7$, that result in a potential with two
length scales, one around $r_{ij}\equiv r_1\approx 1.2 \sigma$ and
the other at  $r_{ij}\equiv r_2 \approx 2 \sigma$~\cite{Oliveira10}. The
potential is shown in the Figure~\ref{fig1}(a). This model does not
exhibit the
directionality or explicitly hydrogen bonds as present in water, however
 it captures the competition between open and close water
tetramers through the competition between the two length scales. As a
result, the pressure versus temperature phase diagram shows a
region where density, diffusion and structural properties are
anomalous in bulk~\cite{Oliveira06a,Oliveira06b,Kell67,Angell76} and
in confined
systems~\cite{Bordin14a,Bordin13a,Bordin14b,Bordin14c,BoK15a,Bordin12b,Krott13a,Krott13b,Krott14a}.

%%%%%%%%%%%%%%%%%%%%%%%%%%%%%%%%%
%%%%%%%%%%%%%%%%%%%%%%%%%%%%

In our system the particles are confined between plates
in $z$-direction. All simulations were done with plates of
size $L_x = L_y = L = 20\sigma$, separated by a fixed
distance $L_z$. A schematic depiction of the system is given
in the Fig.~\ref{fig1} (b). The interaction between the fluid particles
and the molecules in the plates
is purely repulsive and it is given by
the Weeks-Chandler-Andersen (WCA)~\cite{WCA71} potential, namely
%%%%%%%%%%%%%%%%%%%%%%%%%%%%%%%%%%%
\begin{equation}
\label{LJCS}
U^{\rm{WCA}}(z_{ij}) = \left\{ \begin{array}{ll}
4\epsilon\left[ \left(\frac{\sigma}{z_{ij}}\right)^{12} -
\left(\frac{\sigma}{z_{ij}}\right)^6 \right] + \epsilon\;, \qquad z_{ij} \le 2^{1/6}\sigma\;, \\
0\;, \qquad \qquad \qquad \qquad \quad z_{ij}  > 2^{1/6}\sigma\;.
\end{array} \right.
\end{equation}
%%%%%%%%%%%%%%%%%%%%%%%%
\noindent where $z_{ij}$ is the distance between the plates
at $j$ position and the $z$-coordinate of the fluid particle $i$.

%%%%%%%%%%%%%%%%%%%%%%%%%%%%
\section{Methods}
%%%%%%%%%%%%%%%%%%%%%%%%%%%%

%%%%%%%%%%%%%%%%%%%%%%%%%%%%
\subsection{The simulation details}
%%%%%%%%%%%%%%%%%%%%%%%%%%%%

The physical quantities are shown  in reduced units~\cite{Allen},
%%%%%%%%%%%%%%%%%%%%%%%%%%%%
\begin{equation}
\label{red1}
r^*\equiv \frac{r}{\sigma}\;\quad \mbox{and}\quad \rho^{*}\equiv
\rho \sigma^{3}\;,
\end{equation}
%%%%%%%%%%%%%%%%%%%%%%%
for distance and density of particles, respectively, and
%%%%%%%%%%%%%%%%%%%%%%%%%
\begin{equation}
\label{rad2}
P_{||}^*\equiv \frac{P_{||} \sigma^{3}}{\epsilon} \quad \mbox{and}\quad
T^{*}\equiv \frac{k_{B}T}{\epsilon}
\end{equation}
%%%%%%%%%%%%%%%%%%%%%%%%%%%%%%%%%%%%%%%%%%%%%
for the pressure and temperature, respectively. The symbol $*$ will
be omitted in order to simplify the discussion of the paper.

The molecular dynamic simulations were performed at NVT-constant. The systems
have $500$ particles confined between two parallel flat plates in
$z$-direction. The plates have thickness of $\sigma$, area of $L^2$ and
 are separated by a fixed distance $L_z$. The value of $L$ was $20$ in
all the simulations and the density of the system was
 changed varying the value of $L_z$, from $4.3$ to $10.0\sigma$.

The repulsive interaction with the plates creates an exclusion
region in the $z$ direction, consequently  the total density is  corrected to
an effective density~\cite{Ku05,kumar07}. Then the
distance between the plates can be approach by
$L_{ze} \approx L_z -\sigma$, resulting in an effective density of
$\rho = N/(L_{ze} L^2)$. In $x$ and $y$ directions periodic boundary
conditions were employed.

The temperature of the systems was fixed using the
Nos\'e-Hoover heat-bath~\cite{Ho85,Ho86} with a
coupling parameter $Q = 2$. Simulations were performed
for the  temperatures
ranging from $T = 0.050$ to $T = 0.450$. This choice of
temperatures were made on basis of the pressure versus temperature phase
diagram of the bulk system. The initial configuration was generated
placing the fluid particles randomly between the plates. The equations of
motion were integrated using the velocity Verlet algorithm, with time step
of $\delta t = 0.001$. The systems were equilibrated with $5\times10^5$
steps followed by $1\times10^8$ steps for the production of the
thermodynamic averages. A particle-particle interaction
cutoff radius of $r_{\rm cut}/\sigma = 3.5$ was used.

%%%%%%%%%%%%%%%%%%%%%%%%%%%%%%%%%%%%%%%%%%%%%
\subsection{The multifractal spectrum}
%%%%%%%%%%%%%%%%%%%%%%%%%%%%%%%%%%%%%%%%%%%%%
We used a standard box counting method, also called the \emph{capacity} of the set~\cite{Barnsley,Ott,Chhabra,Chhabra-1}, to calculate the fractal dimension on the two-dimension configurational structure of each final equilibrium state, on the layer of the fluid confined. The images are sets of two-dimensional final stages of the simulation process. Each one present a particle configuration for fix densities and diverse values of temperature. The procedure to measure the fractal dimension is the follow: the image is binarized by a high contrast treatment, leaving the particles black and the space between them white. A grid of four random positions cover the entire image with a decreasing size of $\varepsilon$ as the length of the box. The scaling law to relate the number of particles and the size of the boxes follow the relation
\begin{equation}
N\sim\varepsilon^{-D_{q}}
\end{equation}
where $\varepsilon$ acquired successively smaller values of length until the minimum value of $\varepsilon_{0}$ and  $N(\varepsilon)$ are the number of cubes required to coverall the set. The fractal dimension by the box counting method is given by
\begin{equation}
D_{q}=\lim_{\varepsilon \rightarrow 0}\frac{\ln N(\varepsilon)}{\ln (\varepsilon_{0}/\varepsilon)}
\end{equation}
To describe all the statistical properties by the local fractal dimension~\cite{Ott,Halsey}, we used the generalized box counting method~\cite{Hentschel,Feigenbaum,Grassberger,Procaccia} defined as
\begin{equation}
D_{q}=\frac{1}{1-q}\lim_{\varepsilon\rightarrow 0}\frac{ln I(q,\varepsilon)}{ln (\varepsilon_{0}/\varepsilon)}
\end{equation}
where
\begin{equation}
I(q,\varepsilon)=\sum_{i=1}^{N(\varepsilon)}[P_{(i,q)}]^{q}
\end{equation}

We used the scaling exponent defined by Halsey et al.~\cite{Halsey} as $P_{i,q}^{q}\sim\varepsilon_{i}^{\alpha q}$ where $\alpha$ can take a width range of values measuring different regions of the set. The spectrum generated by an infinite set of dimension $D_{q}=D_{0},D_{1},D_{2},...$ measure the scaling structure as a function of the local pattern density. If \emph{q=0} the generalized fractal dimension represent the classic fractal dimension, it means that $D_{f}=D_{q=0}$~\cite{Mureika}. As the image is divided into pieces of size $\varepsilon$, it suggested that the number of times that $\alpha$ in $P_{i,q}$ takes a value between $\alpha'$ and $d\alpha'$  defined as $d\alpha'\rho(\alpha')\varepsilon^{-f(\alpha')}$ where $f(\alpha')$ is a continuous function.
As \emph{q} represents different scaling indices, we can define
%%%%%%%%%%%%%%%%%%%%%%%%%%%%%%%%%%
\begin{equation}
I(q,\varepsilon)=\sum_{i=1}^{N(\varepsilon)}[P_{(i,q)}]^{q}=\int d\alpha'\rho(\alpha')\varepsilon^{-f(\alpha')+q\alpha'}
\end{equation}
%%%%%%%%%%%%%%%%%%%%%%%%%%%%%%%%%%
$\alpha_{i}$ is the  Lipschitz-H\"{o}lder exponent, that characterizes the singularity strength in the \emph{ith} box. The factor $\alpha_{i}$ allows to
quantify the distribution of complexity in an spatial location. The multifractal is a set of overlapping self-similar configurations. In that way, we used the scaling relationship taking into account $f(\alpha)$ as a function to cover a length scales of observations. Defining the number of boxes as a function of the Lipschitz-H\"{o}lder exponent $N(\alpha)$, can be related to the box size $\varepsilon$ as
%%%%%%%%%%%%%%%%%%%%%%%%%%%%%%%%%%
\begin{equation}
N(\alpha)\sim\varepsilon^{-f(\alpha)}
\end{equation}
%%%%%%%%%%%%%%%%%%%%%%%%%%%%%%%%%%

The multi-fractal spectrum show a line of consecutive points for $Q\geq0$ that start on the left side of the spectrum climbing up to the maximum value. Then the values for $Q\leq0$, represented in the spectrum for a dotted line, on the right side begins to descend. The maximum value corresponds to $Q=0$, which is equal to the box counting dimension. To obtain the multi-fractal spectrum we use the plugin \emph{FracLac} for ImageJ\cite{PDF}. Basically $D_{q}$ is the variation of mass as a function of $\varepsilon$ in the image, and give us the behavior as a power series of $\varepsilon$ sizes distorting them by an exponent \emph{q}. We select the case of $D_{f}=D_{q=0}$ as the parameter of order in the images. In the plugin we select four grid positions that cover the total image, and the mode scaled series was selected to see the singularity spectrum results. The final configuration of the simulations, present a particle distribution of particles in black. The parameters of the program were calibrate for this kind of images.

%%%%%%%%%%%%%%%%%%%%%%%%%%%%%%%%%%%%%%%%%%%%%
\section{Results}
\label{}

The temperature versus
density phase diagram of this system was obtained by molecular dynamic
simulations~\cite{Bordin14a} showing the various two
dimensional  phases
present at the  contact layer.

The existence of  phase transitions in the analysis of Bordin et al.~\cite{Bordin14a} was
obtained by computing the
 density versus pressure
isochores and observing instabilities related
to first-order phase transitions.
The general classification employed for by Bordin et al~\cite{Bordin14a}
for defining the different phases  of this system took into account: the
radial distribution function, the particles mobility and a
snapshot of the system. It is interesting
to observe that large structural changes
can occur without phase transition. In order
to understand how very different structures
are connected here we explore the fractal dimension
of each one of these structures.

\begin{figure}[!htb]
\centering
 \begin{tabular}{ccc}
     \includegraphics[width=0.5\textwidth]{rho_x_T_high.eps}
             \includegraphics[width=0.4\textwidth]{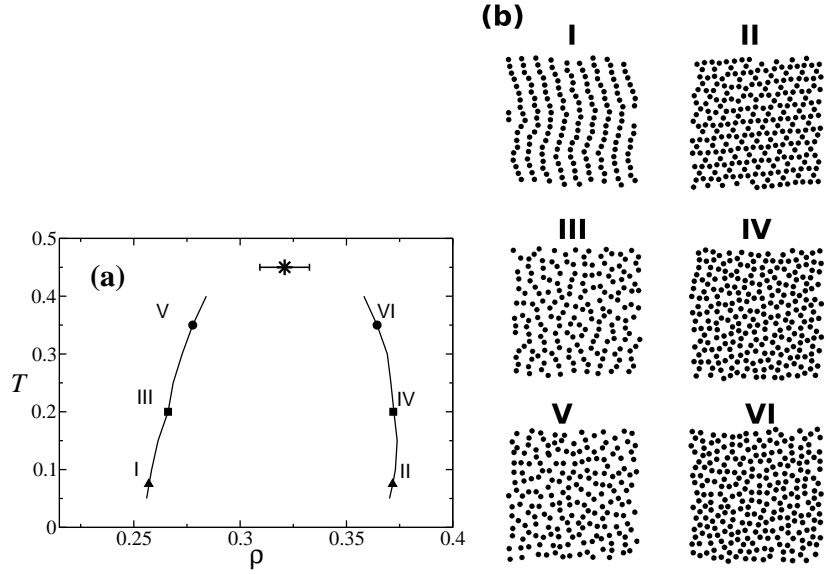}\\
  \tabularnewline
 \end{tabular}\par
  \caption{(a) Temperature versus total density phase diagram for
high densities. The star represents the
  critical point at $T_c = 0.450$ and $\rho_c = 0.321$. The solid
lines delimitate the coexistence
  region between liquid-crystal and hexagonal solid phases. The
triangles, squares and circles are isochores at $T = 0.075$, $0.200$ and
$0.350$, respectively, and
densities $\rho_{I,III,V} \approx 0.260$,
  $\rho_{II,IV,VI} \approx 0.370$. The
corresponding snapshots of the
  contact layer are shown in (b). More details in the text.}\label{rhoTh}
\end{figure}

Now let us explore each density region of the phase diagram.
Figure~\ref{rhoTh} (a) illustrates the temperature versus total
density phase diagram showing the different phases
present at the fluid contact layer~\cite{Bordin14a}.
Figure~\ref{rhoTh} (b) show selected
snapshots of the different phases.
At this high density region  at
low temperatures a liquid-crystal phase (I) coexists with a solid hexagonal
phase (II). At higher values of the temperature both
phases become liquid and increasingly disordered
and for $\rho_{c3}=0.321$ and $T_{c3}=0.45$ the coexistence
disappears at critical point. It is important to notice
that no phase transition is observed between the
liquid-crystal phase: phase I, and the liquid phases: phases III and V. Also no phase transition is
observed between the solid hexagonal
phase, phase II, and the other two liquid
phases, phases IV and VI.

Here we complement this analysis
by identifying the different phases with the correlation between
each final equilibrium state of the phase transition and the degree of ordering.

%%%%%%%%%%%%%%%%%%%%%%%%%%%%%%%%%%%%%%%%%%%%%%%%%%%%%%%
%%%%%%%%%%%%%%%%%%%%%%%%%%%%%%%%%%%%%%%%%%%%%%%%%%%%%%%

The starting point for our analysis are the snapshots of the
system. Fig.~\ref{rhoTh} (b) shows the pictures for  the temperatures
$T = 0.075$ (points I and II), $T=0.200$ (points III and IV) and
$T=0.350$ (points V and VI). These figures represent the final
equilibrium state in which the specific
structure exists. To make more evident this evolution, we
calculated the fractal dimension of each layer at those
different densities and temperatures. This order parameter
gives a quantitative classification for these phase transitions.
%%%%%%%%%%%%%%%%%%%%%%%%%%%%%%%%%%%%%%%%%%%%%%%%%%%%%%%
\begin{figure}[!htb]
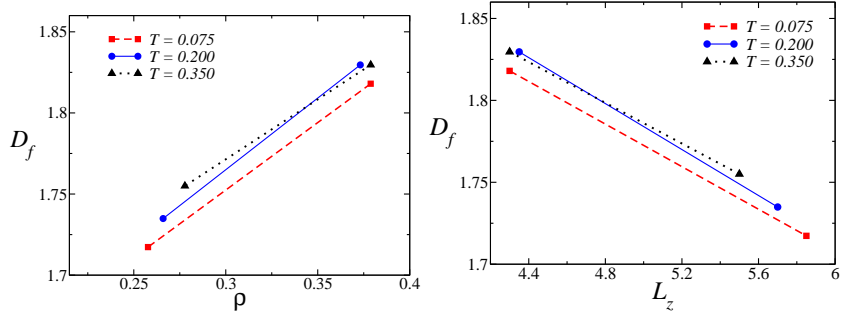

\centering
 \begin{tabular}{ccc}
  \includegraphics[width=0.45\textwidth]{high_density.eps}
  \includegraphics[width=0.45\textwidth]{high_density_Lz.eps}
  \tabularnewline
 \end{tabular}
  \caption{(a) Fractal dimension as function of (a) total density and (b) separation of plates for
  transition at high densities.}\label{DH}
\end{figure}
%%%%%%%%%%%%%%%%%%%%%%%%%%%%%%%%%%%%%%%%%%%%%%%%%%%%%%%

Figure~\ref{DH}(a) presents the maximum values obtained in the singularity spectrum
 as a function of total density, $\rho$, while figure~\ref{DH}(b) shows
it as a function of the  separation between plates, $L_{z}$, for the
 three values of the temperature. In all the cases the fractal dimension
increases with the increasing of confinement. Since the fractal dimension
can be used as measure of the degree of order~\cite{Vicsek,Calleja1}, our results
confirm that for each temperatures the order increases with density. It
is interesting to notice that, the slopes of the constant temperature lines
 are almost the same for all the temperatures analyzed, indicating that
even though the structures change with temperature, the difference in
order between the low density and high density structures, change
proportionally. In addition, the values of the fractal
dimension of the phases I, III and V are very similar, while
the values for phases II, IV and VI are almost the same. This
could explain why  the transformation between them requires
no phase transition.

%%%%%%%%%%%%%%%%%%%%%%%%%
\begin{figure}[!htb]
\centering
 \begin{tabular}{ccc}
    \includegraphics[width=0.5\textwidth]{rho_x_T_mid.eps}
            \includegraphics[width=0.4\textwidth]{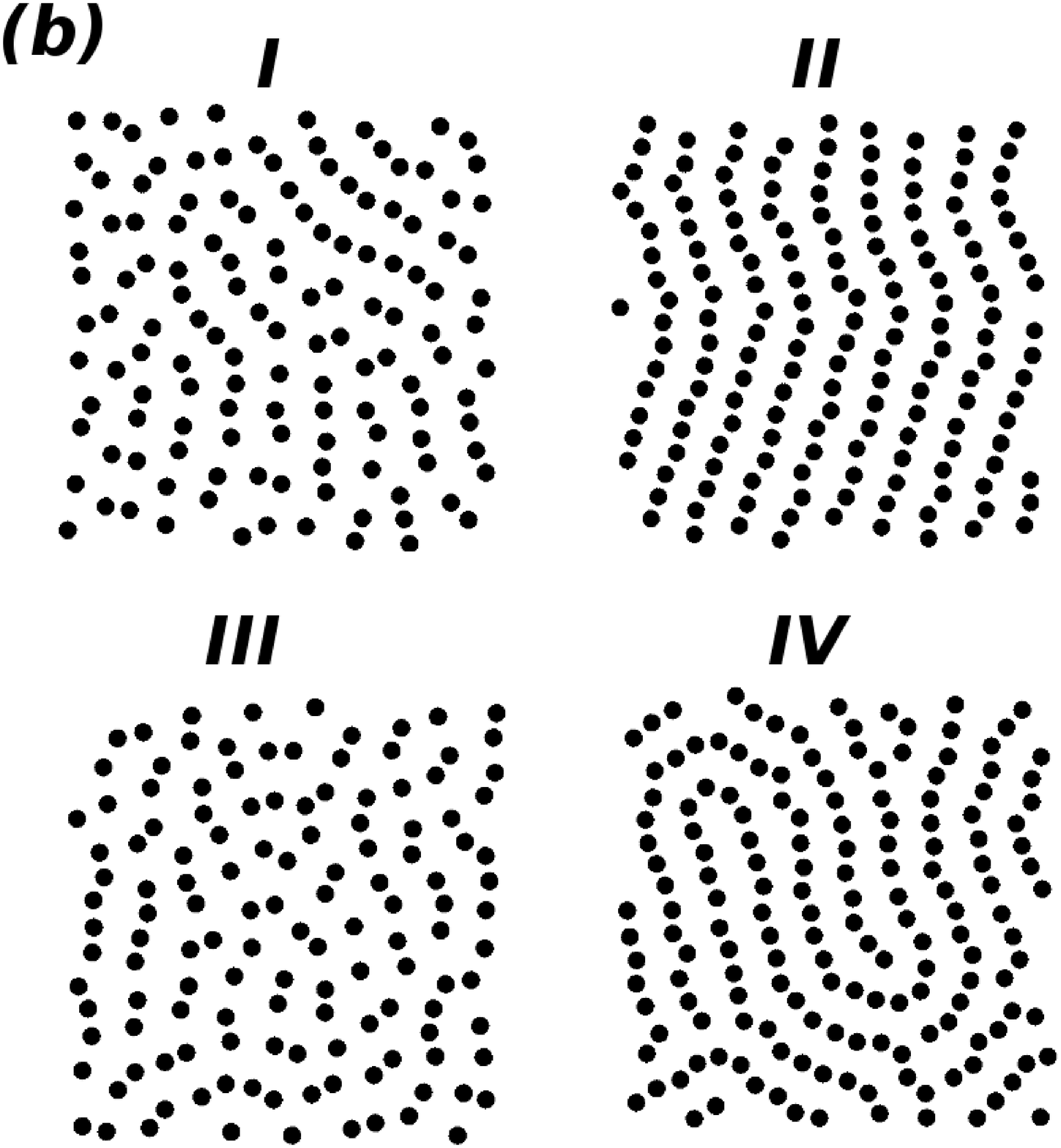}\\
  \tabularnewline
 \end{tabular}\par
  \caption{(a) Temperature versus total density phase diagram for intermediate densities. The star represents the
  critical point at $T_c = 0.20$ and $\rho_c = 0.212$. The solid lines delimitate the coexistence
  region between dimeric liquid and liquid-crystal phases. The snapshots of the contact layer
  are shown in (b) for the points I and II at $T = 0.075$ and for points III and IV at $T = 0.100$, with densities
  $\rho_{I,III} \approx 0.185$ and $\rho_{II,IV} \approx 0.228$.}\label{rhoTm}
\end{figure}
%%%%%%%%%%%%%%%%%%%%%%%%%

Fig.~\ref{rhoTm}(a) illustrates the temperature versus density phase
diagram of the liquid at the contact layer of
the confined system in a region of intermediate
densities~\cite{Bordin14a}. At low temperatures a liquid-crystal
phase (II) coexists with a structured liquid phase (I) made of dimers. When
 the temperature is raised both the liquid-crystal and the structured dimeric liquid becomes disordered, phases (IV) and
(III) in the Fig.~\ref{rhoTm}(b) respectively. As the temperature
is increased even further the coexistence ends at a critical
point at $T_{c2}=0.20$ and $\rho_{c2}=0.212$. It is important
to notice that even though quite different in structure
it is possible to go from the phase I to phase III and
from phase II to phase IV with no phase transition.

The snapshots of the final states of the system at the contact layer
for this intermediate range of densities are depicted in the
Figure~\ref{rhoTm}(b).
The figure illustrates the states I and II at $T = 0.075$ and for
III and IV $T=0.100$, with densities $\rho_{I,III} \approx 0.185$
and $\rho_{II,IV} \approx 0.228$, respectively.
%%%%%%%%%%%%%%%%%%%%%%%%%%%
\begin{figure}[!htb]
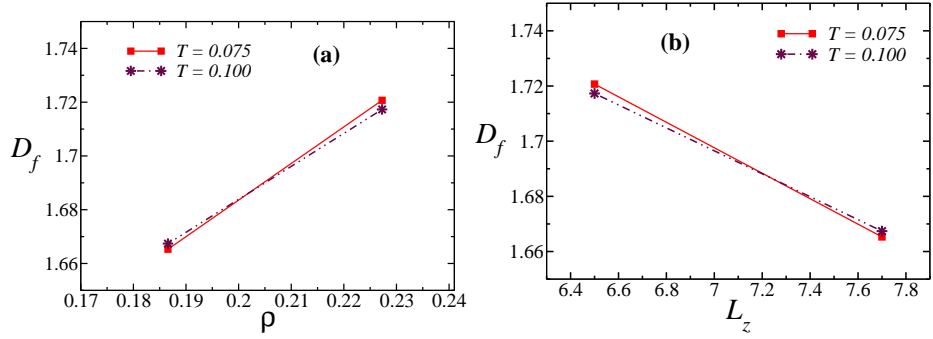

\centering
 \begin{tabular}{ccc}
  \includegraphics[width=0.5\textwidth]{medium_density.eps}
  \includegraphics[width=0.5\textwidth]{medium_density_lz.eps}\\
  \tabularnewline
 \end{tabular}\par
  \caption{(a) Fractal dimension as function of (a) total density and (b) separation of plates for
  transition at intermediate densities.}\label{DM}
\end{figure}
%%%%%%%%%%%%%%%%%%%%%%%%%%%%%%%%%%%%%%%%%%%%%%%%%
Employing these snapshots, the fractal dimensions of these configurations
were computed. The Figure~\ref{DM} shows the degree of ordering
evolution as a function
of total density and of the distance $L_{z}$ for this intermediate
region of densities. Similarly to what happens at high densities
shown in Fig.~\ref{DH} the increase in the confinement and the
decrease of the distance between the wall leads to an increase in the order.
The phases I and III show almost the same
value for the fractal dimension. The same occurs
for the phases II and IV. This result support the
idea that the transformation from one phase to
another structurally very different
with no phase transition might be related to our
findings that they share the same fractal structure.

%%%%%%%%%%%%%%%%%%%%%%%%%%%%%%%%%%%%%%%%%%%%%%%%%
\begin{figure}[!htb]
\centering
 \begin{tabular}{ccc}
   \includegraphics[width=0.5\textwidth]{rho_x_T_low.eps}
  \includegraphics[width=0.5\textwidth]{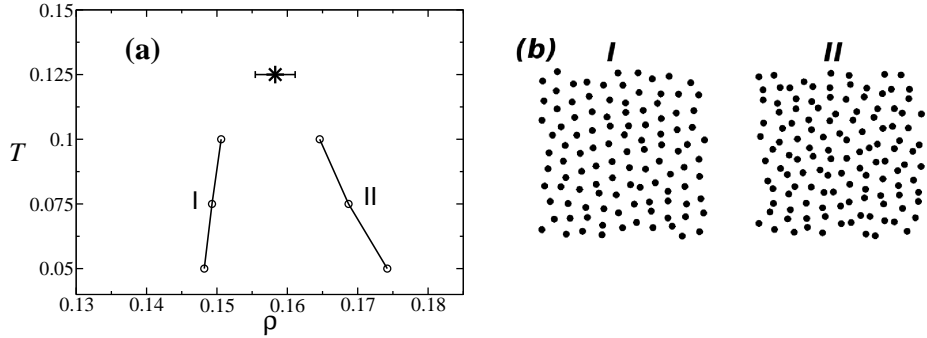}\\

  \tabularnewline
 \end{tabular}\par
  \caption{(a) Temperature versus total density phase diagram for low densities. The star represents the
  critical point at $T_c = 0.125$ and $\rho_c = 0.158$. The solid lines delimitate the coexistence
  region between disordered liquid and dimeric liquid phases. The snapshots of the contact layer
  are shown in (b) for the points I and II at $T = 0.075$, with densities
  $\rho_{I} \approx 0.149$ and $\rho_{II} \approx 0.169$.}\label{rhoTl}
\end{figure}
%%%%%%%%%%%%%%%%%%%%%%%%%%%%%%%%%%%%%%%%%%%%%%%%%

%%%%%%%%%%%%%%%%%%%%%%%%%%%%%%%%%%%%%%%%%%%%%%%%%

\begin{figure}[!htb]
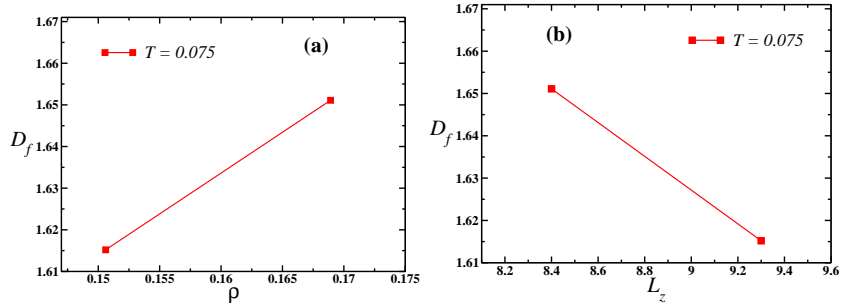

\centering
 \begin{tabular}{ccc}

     \includegraphics[width=0.45\textwidth]{low_density.eps}
  \includegraphics[width=0.45\textwidth]{low_density_lz.eps}
  \tabularnewline
 \end{tabular}\par
  \caption{ Fractal dimension as
  function of (a) density for $T = 0.075$ and
(b) separation between plates.}\label{DL}
\end{figure}
%%%%%%%%%%%%%%%%%%%%%%%%%%%%%%%%%%%%%%%%%%%%%%%%%

Figure~\ref{rhoTl}(a) shows the temperature versus density phase diagram
of the liquid at the contact layer of the confined system in a region
of low densities~\cite{Bordin14a} showing liquid configurations at
coexistence. The coexistence between the two liquid phases ends at a
critical point at $T_{c1}=0.125$ and $\rho_{c1}= 0.158$. The snapshots
of representative phases at the contact layer are shown in the
Figure~\ref{rhoTl}(b). Phases \emph{I} and \emph{II}, with $T = 0.075$
and densities $\rho_{I} \approx 0.149$ and $\rho_{II} \approx 0.169$
are illustrated, respectively.

These equilibrium configurations were employed to compute the
fractal dimension of the system. In the Fig.~\ref{DL}(a), we
show the fractal dimension as function of density, while in the
Figure~\ref{DL} (b) the fractal dimension is computed as a
function of the separation between plates. At this temperature, the
disorder degree varies from $D_{f}=1.6152$ ($\rho \approx 0.150$)
to $D_{f}=1.6511$ ($\rho \approx 0.170$). Even all the
configurations being liquid states, we identify different
degree of disorder in each case.
%%%%%%%%%%%%%%%%%%%%%%%%%%%%%%%%%%%%%%%%%
\begin{figure}[h!]
\centering
 \begin{tabular}{ccc}
  \includegraphics[width=0.5\textwidth]{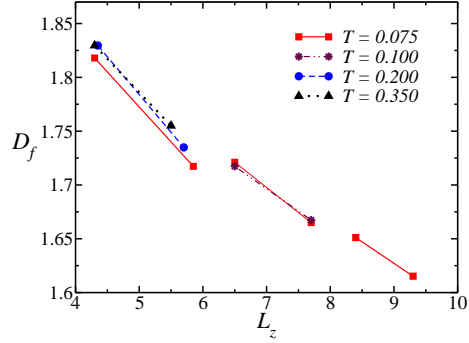}
  \tabularnewline
 \end{tabular}\par
  \caption{Order evolution as function of $L_z$ for all the cases studied.}\label{allF}
\end{figure}
%%%%%%%%%%%%%%%%%%%%%%%%%%%%%%%%%%%%%%%%%

In all the densities analyzed the fractal analysis confirms the
results obtained with molecular dynamic simulations. Now in order
to check if the different phase transitions show drastic differences
in the evolution of the fractal dimension, the degree of ordering is
compared. Figure~\ref{allF} illustrates the fractal dimension as a function
 of the distance between the plates, $L_{z}$ for all the cases studied
here. The fractal dimension not only increases with the density but
also a drastic change of slope is observed at very high densities. This
region of pressures and temperatures represents the solid to liquid phase
transitions where a high degree of ordering is expected confirming the
suggestions of the molecular dynamic simulations.

%%%%%%%%%%%%%%%%%%%%%%%%%%%%%%%%%%%%%%%%%
\section{Conclusion}
%%%%%%%%%%%%%%%%%%%%%%%%%%%%%%%%%%%%%%%%%

In this work we have
explored the use of the  fractal dimension to
analyze the phases present
in an anomalous fluid confined by repulsive
walls.

 The degree of configurational order-disorder of the
confined liquid was analyzed using the fractal spectrum approach
through image analysis. We found that different phases
separated by phase transitions show
a very different fractal dimension that
increases with the increasing order
of the structure of the phase. Complementary
to this result we also found that
phases that are not
separated by phase transitions show
a very similar fractal dimension.
This result shade some light in
the odd possibility of a continuous
transition between two structurally very
different phases.

%%%%%%%%%%%%%%%%%%%%%%%%%%%%%
\section{Acknowledgment}
%%%%%%%%%%%%%%%%%%%%%%%%%%%%%

We thank the Brazilian agencies CNPq, INCT-FCx, and Capes for
 the financial support. E.M.D.C. Bolsista do CNPq - Brazil.

%%%%%%%%%%%%%%%%%%%%%%%%%%%%%

%%%%%%%%%%%%%%%%%%%%%%%%%%%%%%%%%%%%%%%%%
\bibliography{mybibfile}

\bibliographystyle{aip}

\end{document}